\documentclass[aps,nofootinbib,prd,eqsecnum,showpacs,showkeys,preprintnumbers]{revtex4-2}
\usepackage[caption=false]{subfig}
\usepackage[utf8]{inputenc}
\usepackage{graphicx}
\usepackage{amsmath}
\usepackage{amsfonts}
\usepackage{amssymb}
\usepackage{float}
\usepackage{color}
\usepackage{bm}
\usepackage{mathrsfs}
\usepackage{epstopdf}
\usepackage{url}
\usepackage{footnote}
\usepackage{textcomp}
\usepackage{ulem}
\usepackage{esint}
\usepackage[unicode=true, pdfusetitle,
 bookmarks=true,bookmarksnumbered=false,bookmarksopen=false,
 breaklinks=false,pdfborder={0 0 1},backref=false,colorlinks=false]{hyperref}
\usepackage{multirow}
\usepackage{pifont}
\usepackage{soul}

\begin{document}

\title{Signatures of the Early Universe: Uncovering Cosmological Footprints}
\author{K. El Bourakadi$^{1,2}$}
\email{k.elbourakadi@yahoo.com}
\author{Z. Sakhi$^{2}$}
\email{zb.sakhi@gmail.com}
\author{M. Bennai$^{2}$}
\email{mdbennai@yahoo.fr}

\begin{abstract}
The post-inflationary epochs are critical for comprehending the early
evolution of our Universe. This article delves into the cosmological
signatures that shed light on these early epochs, particularly focusing on
the generation of various phenomena such as matter production via inflaton
oscillation and parametric resonance, primordial black holes, and
gravitational waves. We review the theoretical frameworks that could produce
these signatures and discuss the current observational constraints along
with prospects for future detection. Furthermore, we explore the
implications of such observations for our understanding of the physics of the early Universe.
\end{abstract}


\affiliation{$^{1}$\small Subatomic Research and Applications Team, Faculty of Science
Ben M'sik,\\
{\small Casablanca Hassan II University, Morocco}\\} 
\affiliation{$^{2}$\small Physics and Quantum Technology Team, LPMC, Ben M’sik Faculty of Sciences, Casablanca Hassan II University, Casablanca, Morocco}
\maketitle

\section{\label{sec:1} Introduction}

The theory of reheating, a crucial yet not fully explored aspect of
inflationary cosmology, has recently attracted significant attention,
particularly regarding the initial phase known as the parametric resonance
phenomenon. In this scenario, the inflaton field rapidly transfers energy to
other interacting scalar and vector fields without reaching thermal
equilibrium, this phenomenon is also known as preheating \cite{I1,K2}. A
comprehensive study of preheating in the chaotic inflation model, which
involves a massive inflaton field $\phi$ interacting with a massless scalar
field $\chi$, revealed that resonance in these models is effective only if
it is very large \cite{I3}. In this context, preheating in an expanding
universe manifests as a stochastic process. It is commonly assumed that
following inflation, the massive inflaton field begins oscillating around
its minimum, resulting in decay processes where the decay products rapidly
reach thermal equilibrium, forming a thermal bath with a final temperature $%
T_{re}$ \cite{I4}. The reheating temperature is typically defined as the
point at which the energy density of the newly formed radiation bath equals
the energy density of the inflaton's oscillations. Studies such in Ref. \cite{I5}
have explored this concept, showing that the behavior of the thermal bath is
influenced by the structure of the potential driving the inflaton's
oscillations. Various methods can be employed to study cosmic reheating,
including direct tests using a stochastic gravitational wave background \cite%
{I6}. Indirect tests through Cosmic Microwave Background (CMB) observations
also play a crucial role, analyzing the impact of the altered equation of
state $\omega$ during the reheating period on the post-inflationary
expansion history of the universe \cite{I7,I8}. Additionally, methods used
in Ref. \cite{I9} can be utilized to validate the reheating temperature $%
T_{re}$, which has been examined across several models, including
single-field inflation \cite{I10,I11}, power law inflation \cite{I12},
natural inflation \cite{I13}, Hilltop inflation \cite{I12,I14}, $\alpha$%
-attractor inflation \cite{I12,K7}, tachyon inflation \cite{I15}, and
various SUSY models \cite{I16}.

The observed cosmological perturbations appear to originate from a primordial curvature perturbation, $\zeta .$ This
perturbation is present when the smallest cosmological scales approach
horizon entry, which occurs at a temperature of approximately $1MeV$ \cite%
{I17}. During this time, the curvature perturbation is time-independent as
the cosmic fluid is predominantly radiation-dominated. The prevailing theory
suggests that the value of $\zeta $ at each position is determined by one or
more scalar fields, evaluated during a specific initial epoch of inflation.
The perturbations in these fields are expected to occur from vacuum
fluctuations. In the initial model, $\zeta $ is entirely generated by the
perturbation of the inflaton field during a slow roll inflation model. In
this scenario, $\zeta $ is established from the beginning and remains
constant thereafter. However, alternative models propose that a significant
contribution to $\zeta $ can arise from the perturbation of additional
fields, other than the inflaton field. Initially, this contribution might be
negligible but grows to its final value later, typically after the end of
inflation. The inhomogeneities in the field result in a curvature
perturbation, which is then transferred to matter and radiation when the
inflaton field decays. In the simplest scenario, the curvature perturbation
remains constant on scales much larger than the Hubble length and
specifically does not change during the inflaton's decay \cite{L8}.

During the preheating phase, the inflaton field rapidly decays, leading to
an exponential increase in particle production. This process continues until
the backreaction from particle production interrupts the parametric resonance,
after which the newly generated particles undergo thermalization, then
followed by a radiation-dominated universe. The resonant amplification of
field fluctuations during preheating may cause an overproduction of
primordial black holes (PBHs) on relatively small scales. These scales exit
the Hubble radius near the end of inflation, The amplified fluctuations can
potentially form PBHs upon re-entering the Hubble radius during the
radiation-dominated era \cite{L7,I18,I19}. The potential observational
imprints of PBHs could address several astrophysical questions. Such
signatures might explain non-linear seeds of large-scale structure, and the
evaporation of PBHs could account for point-like gamma-ray sources \cite{I20}%
.

According to general relativity, the present-day universe is expected to
contain a pervasive gravitational wave background (GWB) originating from
various sources. These include relic stochastic backgrounds from the early
universe, phase transitions, inflation, turbulent plasmas, and cosmic
strings \cite{I21}. In fact, confirming the inflationary paradigm is closely
tied to the detection of relic gravitational waves (GWs), which are regarded
as a definitive signal. Recent advances in observational cosmology have
shown significant progress in this area. Various collaborations, including
NANOGrav \cite{I23,I24,I25,I26}, the European PTA (EPTA)/Indian PTA (InPTA) 
\cite{I27,I28}, and the Parkes PTA (PPTA) \cite{I29}, have provided
compelling evidence for the existence of a Stochastic Gravitational Wave
Background (SGWB) in the nanohertz (nHz) frequency range. While these
findings are mainly attributed to astrophysical sources, it is essential to
explore their potential cosmological origins. The paper is organized as
follows. In Sect. \ref{sec:2}, we give an overview of Cosmological
inflation. In Sect. \ref{sec:3}, we provide a summary of the
post-inflationary Universe. We further perform an analysis on the formalism
of inflaton oscillations and reheating in Sect. \ref{sec:4}. In Sect. \ref%
{sec:5}, we discuss the preheating scenario. In Sect. \ref{sec:6}, we
discuss the cosmological signatures. We conclude in Sect. \ref{sec:7}.

\section{\label{sec:2}Overview of Inflationary Cosmology}

\subsection{The Standard Cosmological Model}

The standard Big Bang cosmology is built upon three key theoretical
foundations: the cosmological principle, Einstein's general theory of
relativity, and a classical description of matter as a perfect fluid. The
cosmological principle postulates that the Universe is homogeneous on a
large scale, allowing the space-time metric to be described by the
Friedmann-Robertson-Walker (FRW) form,

\begin{equation}
ds^{2}=-dt^{2}+a(t)^{2}\left[ \frac{dr^{2}}{1-kr^{2}}+r^{2}\left( d\theta
^{2}+\sin ^{2}\theta d\varphi ^{2}\right) \right] .  \label{eq1}
\end{equation}%
Here, $k$ defines the nature of spacial curvature.
Typically, we assume $k=0$, which indicates a spatially flat Universe. For a homogeneous and isotropic Universe with a zero cosmological constant, the Einstein equations FRW equations,%
\begin{eqnarray}
\left( \frac{\dot{a}}{a}\right) ^{2} &=&\frac{8\pi G}{3}\rho +\frac{k}{a^{2}}%
,  \label{eq2} \\
\frac{\ddot{a}}{a} &=&-\frac{4\pi G}{3}\left( \rho +3p\right) .  \label{eq3}
\end{eqnarray}%
By combining these equations, the continuity equation can be derived as,%
\begin{equation}
\dot{\rho}=-3H\left( \rho +p\right) .  \label{eq4}
\end{equation}%
Another fundamental concept of standard cosmology is that matter can be
modeled as a classical ideal gas that adheres to a specific equation of
state $p=\omega \rho .$ For cold matter (dust), the pressure can be
neglected, resulting in $\omega =0$. Consequently, equation (\ref{eq4})
simplifies to:%
\begin{equation}
\rho _{m}(t)\sim a^{-3}(t).  \label{eq5}
\end{equation}%
The equation above pertains to the energy density of cold matter, denoted by 
$\rho _{m}$. In contrast, for radiation, the equation of state is $\omega
=1/3$, and substituting this value into equation (\ref{eq4}) results in:%
\begin{equation}
\rho _{r}(t)\sim a^{-4}(t),  \label{eq6}
\end{equation}%
here, $\rho _{r}(t)$ represents the energy density of radiation. The
standard Big Bang cosmology model is crucial for understanding the evolution
and the large-scale structure of the Universe. It provides a framework for
explaining the observed homogeneity, isotropy, and the relationship between
its matter content and geometry. In fact, FRW equations can predict the
Universe's behavior over time, including the formation of structures like
galaxies and the cosmic microwave background radiation. This model also
serves as a foundation for exploring more complex phenomena, such as dark
matter, dark energy, and the potential for various topologies and geometries
in the Universe.

\subsection{The Inflationary model}

Several critical issues are associated with standard Big Bang cosmology,
though none directly contradict observations, such as the homogeneity and
the flatness problems. The concept of inflation addresses these issues and
is relatively straightforward \cite{G1,G2,G3,G4,G5}. The underlying
assumption is that there is a period, beginning at $t_{i}$ and ending at $%
t_{re}$ (referred to as the "reheating time"), during which the The universe
experiences exponential expansion, characterized by:%
\begin{equation}
a(t)\sim e^{Ht},  \label{eq7}
\end{equation}%
with $t\in \left[ t_{i},t_{re}\right] .$\ This period, known as the "de
Sitter" or "inflationary" phase, and it is defined by a Hubble constant $H$.
The success of Big Bang nucleosynthesis sets an upper limit for the
reheating time $\left( t_{re}\right) $ as,%
\begin{equation}
t_{re}\ll t_{NS},  \label{eq8}
\end{equation}%
here, $t_{NS}$ refers to the time of nucleosynthesis. The most significant
advantage of inflation is that it offers a causal mechanism for creating the
primordial perturbations necessary for the formation of galaxies, clusters,
and other large-scale structures. These perturbations are believed to arise
from a causal microphysical process, which can only have a coherent effect
on length scales smaller than the Hubble radius $\ell _{H}(t)=H^{-1}(t)$.
The Hubble radius $\ell _{H}(t)$ can be understood as the maximum distance
that light (and thus any causal effects) can travel within the expansion.

The crucial question is how to achieve inflation. According to the FRW
equations, an exponential expansion of the scale factor can only occur if
the equation of state of matter satisfies $p=-\rho $. However, this equation
of state is incompatible with the conventional cosmological model's
depiction of matter as an ideal gas of classical particles. Consequently,
the ideal gas description of matter breaks down in the early Universe.
Instead, matter must be characterized using quantum field theory (QFT). In
this framework, classical general relativity describes the evolution of
space and time in our Universe, while QFT describes its matter content. This
combined approach allows for the realization of inflation.

\section{\label{sec:3}Overview of the Post-Inflationary Universe}

\subsection{What Occurs After Inflation Ends?}

After Big Bang Nucleosynthesis (BBN), the universe may experience various
epochs, potentially involving modifications to the standard cosmological
model or the introduction of new components. These post-inflationary periods
can be categorized into different phases, such as the preheating and
thermalization era, which begins at the end of inflation and transitions
into the standard radiation-dominated (RD) era, either abruptly or
gradually. Additionally, other epochs may follow the RD era.

Once the inflationary period is ended, the energy density of the inflaton
field ($\phi$) must be transferred to Standard Model (SM) particles and Dark
Matter (DM) to enter the standard RD era. This intermediate period, known as
preheating and reheating \cite{G6,G7}, depends on the behavior of the inflaton
potential near its minimum and the interactions between the inflaton field
and other fields. Reheating can involve both perturbative and
non-perturbative processes. During the oscillations of the condensate, the
time-varying effective inflaton mass allows energy to be resonantly
transferred from the condensate to shorter wavelength modes, resulting in
the rapid and non-adiabatic amplification of fluctuations with short
wavelengths \cite{G19}.

A possible extension to this scenario involves exploring the interactions
between the inflaton field and potential new physics beyond the Standard
Model. These interactions could lead to the production of exotic particles
or the manifestation of new forces, which might leave imprints on the CMB or
influence the formation of large-scale structures. Additionally, the study
of non-standard reheating mechanisms, such as those involving axion-like
particles or dark sector dynamics could provide further insights into the
the early universe and help reconcile any discrepancies between theoretical
predictions and observational data.

\subsection{Later Eras with with $\protect\omega \geq 0$}

The reheating process following inflation is highly model-dependent,
especially in multi-field inflation scenarios, which can lead to the
occurrence of an epoch after reheating where the equation of state is $\omega \geq 0$. The duration of
these post-reheating eras depends on the shape of the inflaton potential. If 
$V(\phi )$ has a quadratic form near the minimum, the epoch following
reheating is a Matter-Dominated (MD) phase with $\omega =0$, and the length
of this phase depends on the gravitational interactions of the inflaton
condensate \cite{G8,G9}. Conversely, if $V(\phi )$ has a quartic shape near
the minimum, the epoch following reheating is a Radiation-Dominated (RD)
phase with $\omega =1/3$ \cite{G10,G11}. However, in practice, the inflaton
must couple to other fields to decay completely after the resonant decay of
the condensate.

In various well-motivated scenarios, a species of particle $\phi $ can
dominate the universe after reheating with a general equation of state (EoS) 
$\omega $, where the energy density $\rho _{\phi }$ scales as $a^{-3\left(
1+\omega \right) }$. In such scenario is the possibility of a
Matter-Dominated (MD) epoch ($\omega =0$) that can arise when a heavy field
drives the energy density of the universe \cite{G12,G13}. A well-supported
example of such an epoch is the domination of moduli fields in various
string inflation models \cite{G14}. Conversely, an epoch dominated by the
kinetic energy density of a rapidly rolling field has an equation of state
close to 1. This occurs after quintessential inflation when the inflaton
field quickly rolls down towards the potential relevant for future dark
energy \cite{G15}. Different
values of $\omega $ can arise when a scalar field oscillates with a specific
potential shape \cite{G16}, or in scenarios such as braneworld cosmologies 
\cite{Gg16} or scalar-tensor theories of gravity \cite{G17}. Furthermore, a
stiff equation of state ($1/3<\omega <1$) may occur when a sterile field
dominates the post-inflationary phase with a significant energy contribution 
\cite{G18}.

For the universe to evolve into a standard radiation-dominated state at
temperature $T_{RD}$ in the epoch where $\omega \geq 0$, two main methods can be
employed. The first method involves the dominant of the field $\phi$ which
must be decaying with a decay rate $\Gamma _{\phi }$, allowing the
relativistic decay products to dominate and the transition to occur which is
described as \cite{G19},

\begin{equation}
\dot{\rho}_{\phi }+3\left( 1+\omega \right) H\rho_{\phi }=-\Gamma _{\phi
}\rho _{\phi },  \label{eq9}
\end{equation}%
where $H=\dot{a}/a$ is the Hubble parameter. The second method occurs when
the equation of state $\omega >1/3$; in this case, the energy density of the
species dilutes more rapidly than that of radiation, allowing radiation to
naturally become dominant.

\subsection{Primordial Perturbations}

The prevailing scenario involves the generation of scalar fluctuations ($%
\zeta $: curvature perturbation) during the primordial epoch of inflation.
In single-field inflation, these fluctuations quickly freeze after exiting
the horizon. In multi-field inflation scenarios, however, these
perturbations continue to grow even in the superhorizon regime until the end
of inflation. These perturbations lead to the formation of classical density
fluctuations, described by:

\begin{equation}
\delta (\mathbf{x},t)=\frac{\rho -\rho _{b}}{\rho _{b}},  \label{eq10}
\end{equation}%
When these perturbations re-enter the horizon after inflation, they result in classical density fluctuations relative to the background energy density $%
\rho _{b}$ which can be given as,

\begin{equation}
\delta (\mathbf{x},t)=\frac{2\left( 1+\omega \right) }{5+3\omega }\left( 
\frac{1}{aH}\right) ^{2}\nabla ^{2}\zeta (\mathbf{x},t),  \label{eq11}
\end{equation}%
The scale factor $a$ and the equation of state $\omega $ of the background
at the time of re-entry determine the classical density fluctuations. The
overdensities that arise from these classical density fluctuations grow
within the post-inflationary horizon, with the growth behavior influenced by
the value of $\omega $. If the mass corresponding to a given scale $\mathbf{R%
}$ exceeds the Jeans mass, the overdense regions on that scale $\mathbf{R}$
will cease expanding and collapse under gravitational pressure.

Numerous studies in the literature investigate inflationary scenarios where
scalar fluctuations undergo significant growth during inflation, leading to
a power spectrum $P_{\zeta }(k)$ that peaks at a specific wavenumber $%
\mathbf{k}_{p}$, comparable to the wavenumber of the CMB fluctuations. In
single-field inflation models, when the inflaton field slows down
considerably in its potential, it can enter an ultra slow-roll (USR) phase.
During this phase, the slow-roll parameters, which indicate the rate of
change of the inflaton field and are given by,%
\begin{eqnarray}
\epsilon &=&\frac{M_{p}^{2}}{2}\left( \frac{V^{\prime }}{V}\right) ^{2},
\label{eq12} \\
\eta &=&M_{p}^{2}\left( \frac{V^{\prime \prime }}{V}\right) .  \label{eq13}
\end{eqnarray}%
To reduce the power spectrum of curvature perturbations $P_{\zeta }$ from
about $10^{-9}$ at Cosmic Microwave Background (CMB) scales to around $%
10^{-2}$ at smaller scales, the potential energy needs to decrease by
approximately $10^{7}$ orders of magnitude.

In multi-field models of inflation, interactions with a secondary field can
significantly increase the power spectrum $P_{\zeta }(k)$, as the inflaton's
velocity is influenced by the entire multi-field potential. An example is
the hybrid inflation model \cite{G20}, where the gentle waterfall phase
amplifies $P_{\zeta }(k)$. There are various ways in which multiple fields
during inflation can cause an increase in the power spectrum \cite{G21}.
Another mechanism involves a large turning rate in the field space or
inducing instabilities in the isocurvature fluctuations that transfer to
curvature fluctuations \cite{G22,G23,G24,G25}. 

In conclusion, both single-field and multi-field inflationary models provide
mechanisms for the generation and amplification of primordial scalar
fluctuations. While single-field models often rely on ultra slow-roll phases
to achieve significant growth in perturbations, multi-field models offer a
variety of interactions that can enhance the power spectrum. These
mechanisms are critical for understanding the formation of large-scale
structures in the universe and offer potential explanations for the observed
characteristics of the cosmic microwave background and other astrophysical
phenomena.

\section{\label{sec:4}Inflaton Oscillations and the Reheating Process}

\subsection{ Oscillations of the Inflaton Field}

According to most inflation models, after the rapid exponential expansion
phase, the inflaton enters a period of oscillations around a minimum point.
These oscillations continue until the inflaton decays, initiating the
reheating process \cite{K0}. In the small coupling limit, the perturbative
reheating approximation is crucial for characterizing the post-inflationary
dynamics. This approximation generally assumes that the inflaton is a
massive field governed by a quadratic potential centered at the origin. For
simplicity, if we assume that the decay of the inflaton occurs through
fermion production ($\phi \rightarrow \bar{\psi}\psi $), the decay rate can
be easily parameterized as,%
\begin{equation}
\Gamma _{\phi }\equiv \frac{y^{2}}{8\pi }m_{\phi }.  \label{eq14}
\end{equation}%
The strength of the decay is determined by the effective Yukawa coupling,
denoted by $y$. Assuming that the decay products of $\phi $ are relativistic
at the moment of their creation and quickly thermalize within a timescale
much shorter than $\Gamma _{\phi }^{-1}$, they form a thermal bath that
eventually results in a radiation-dominated universe once the energy density
of $\phi $ is fully depleted. The highest temperature achieved by this
plasma after the inflaton decay is known as the reheating temperature,
typically expressed as a generic parameter \cite{K1},%
\begin{equation}
T_{re}\sim \left( \Gamma _{\phi }M_{p}\right) ^{1/2}.  \label{eq15}
\end{equation}%
During the reheating phase following the exponential expansion of inflation,
the inflaton field will undergo damped oscillations around $\phi =0$.
Ignoring decay effects for the moment, the equation of motion governing the
behavior of $\phi $ is expressed as:%
\begin{equation}
\ddot{\phi}+3H\dot{\phi}+V^{\prime }(\phi )=0,  \label{eq16}
\end{equation}%
while the energy density and pressure within the scalar field are written as
follows,%
\begin{equation}
\rho _{\phi }=\frac{1}{2}\dot{\phi}^{2}+V(\phi );~p_{\phi }=\frac{1}{2}\dot{%
\phi}^{2}-V(\phi ).  \label{eq17}
\end{equation}%
Moreover, the continuity equation can be expressed as: 
\begin{equation}
\dot{\rho}_{\phi }+3H\left( \rho _{\phi }+~p_{\phi }\right) =0.  \label{eq18}
\end{equation}

The solution of equation (\ref{eq16}) that describes the temporal behavior
of a quadratic form of inflation $V(\phi )=\frac{1}{2}m^{2}\phi ^{2}$
following the inflation phase can be given as \cite{K2}: 
\begin{eqnarray}\phi (t)=\Phi (t)\cdot \sin (mt),  \label{eq19}\\
\Phi (t) \approx \frac{M_{p}}{3mt} \approx  \frac{M_{p}}{20N}. \label{eq19b}
\end{eqnarray}%
Here $\Phi (t)$ is the amplitude of oscillations, and $N$ is the number of oscillations since the end of inflation. Furthermore, the effects of inflaton decay governing the dynamics of $\phi $%
\ can be described as follows :%
\begin{equation}
\ddot{\phi}+\left( 3H+\Gamma _{\phi }\right) \dot{\phi}+V^{\prime }(\phi ) =0.
\end{equation}

\begin{figure}[tbp]
\centering
\includegraphics[width=14cm]{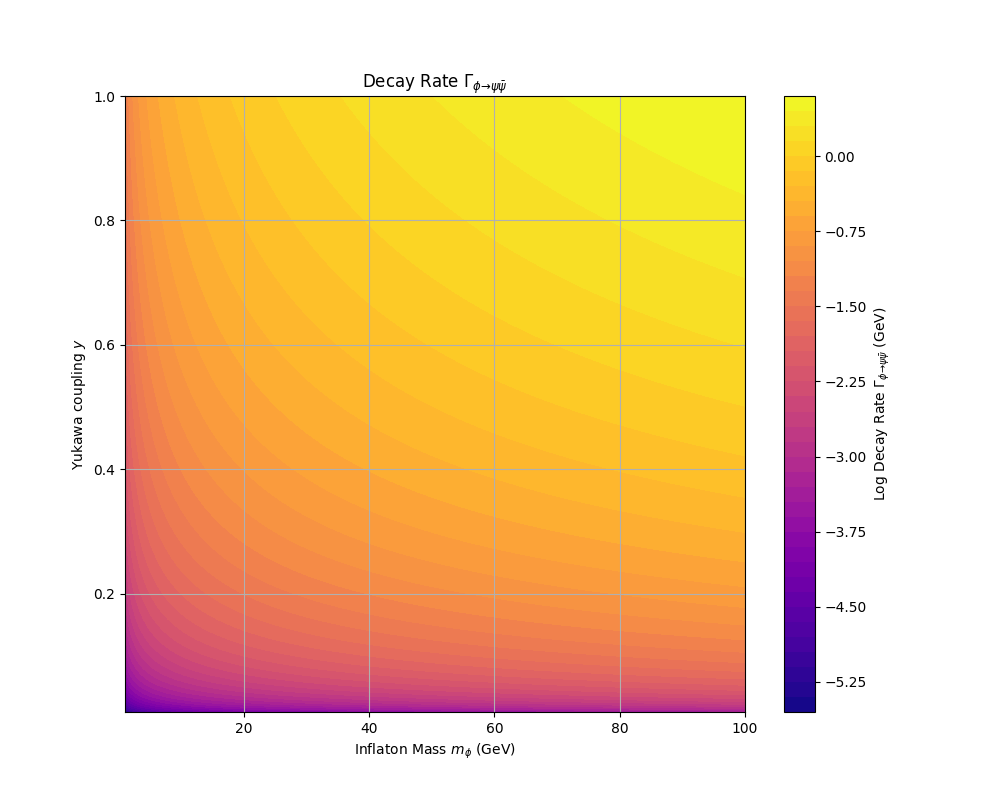}
\caption{{}The Decay Rate of  $\phi$ as a Function of Inflaton Mass and Yukawa Coupling.}
\label{fig:1}
\end{figure}

Fig. \ref{fig:1} presents a heat map illustrating the logarithmic decay rate
of the inflaton field $\left( \phi \right) $\ into a fermion-antifermion
pair $\Gamma _{\phi \rightarrow \psi \bar{\psi}} $ plotted as a function of
the inflaton mass $\left( m_{\phi }\right) $\ and the effective Yukawa
coupling $\left( y\right) $. The x-axis represents $m_{\phi } $ in the range 
$\left( 1GeV\right) $\ to $\left( 100GeV\right) $, while the y-axis
represents the effective Yukawa coupling taking values around $\left[ 0.1,1%
\right] .$ $\Gamma _{\phi \rightarrow \psi \bar{\psi}} $ is computed using
the formula in Eq. \ref{eq14}, showing that the inflaton mass and the Yukawa
coupling significantly influence the decay rate. The plot reveals that larger values of the inflaton mass and the Yukawa coupling lead to increased values of $\Gamma _{\phi \rightarrow \psi \bar{\psi}} $. This dependency helps understanding particle production mechanisms in the early
universe and the subsequent thermalization processes. Understanding the rate $\Gamma _{\phi \rightarrow \psi 
\bar{\psi}}$ behavior is fundamental for modeling the energy transfer from
the inflaton field to standard model particles, which determines the
efficiency of reheating and sets initial conditions for the subsequent
phases of cosmic evolution. \ 

\subsection{Inflaton Decay Mechanism}

When the inflaton field become coupled to Standard Model fields or dark
matter, its oscillations are significantly damped due to the decay process.
To maintain a general perspective, we consider the subsequent potential
contributions to the Lagrangian that can occur from various interaction
terms, such as Yukawa couplings, gauge interactions, or higher-dimensional
operators. By integrating these interactions into the Lagrangian, we can
determine the decay channels and rates of the inflaton field as follows
\cite{K3}, %
\begin{equation}
L\supset 
\begin{cases}
h\phi \bar{\psi}\psi ~\ \ \phi \rightarrow \bar{\psi}\psi  \\ 
\mu \phi \chi \chi ~\ \ \phi \rightarrow \chi \chi  \\ 
g\phi ^{2}\chi ^{2}~\ \ \phi \phi \rightarrow \chi \chi 
\end{cases}%
,  \label{eq23}
\end{equation}%
In this context, $\psi $\ represents fermions and $\chi $\ denotes bosons.
The parameters we are considering include the dimensionless Yukawa-like
coupling, $h$, the four-point coupling, $g$, and the dimensionful coupling, $%
\mu $. It is important to note that, while our investigation is focused on
these three specific cases, our methodology can be readily extended to cover
more unconventional inflaton-matter couplings. We will now analyze the decay
channel \cite{K3},%
\begin{eqnarray}
\Gamma _{\phi \rightarrow \bar{\psi}\psi } &\equiv &\frac{h_{eff}^{2}}{8\pi }%
m_{\phi },  \label{eq24} \\
\Gamma _{\phi \rightarrow \chi \chi } &\equiv &\frac{\mu _{eff}^{2}}{8\pi
m_{\phi }},  \label{eq25} \\
\Gamma _{\phi \phi \rightarrow \chi \chi } &\equiv &\frac{g_{eff}^{2}}{8\pi }%
\frac{\rho _{\phi }}{m_{\phi }^{3}},  \label{eq26}
\end{eqnarray}%
here, we have introduced the effective couplings $h_{eff},\mu _{eff}$\ and $%
g_{eff}.$ By analyzing these expressions, it becomes clear how the form of
the inflaton potential influences its decay rate through its mass, $m_{\phi }
$, and its density, $\rho _{\phi }$. The fundamental concept of reheating
and inflaton oscillation, as previously discussed, is both simple and
intuitive. This approach has proven to be a highly effective way to describe
reheating after inflation across a wide range of realistic inflationary
models.

\begin{figure}[tbp]
\centering
\includegraphics[width=18cm]{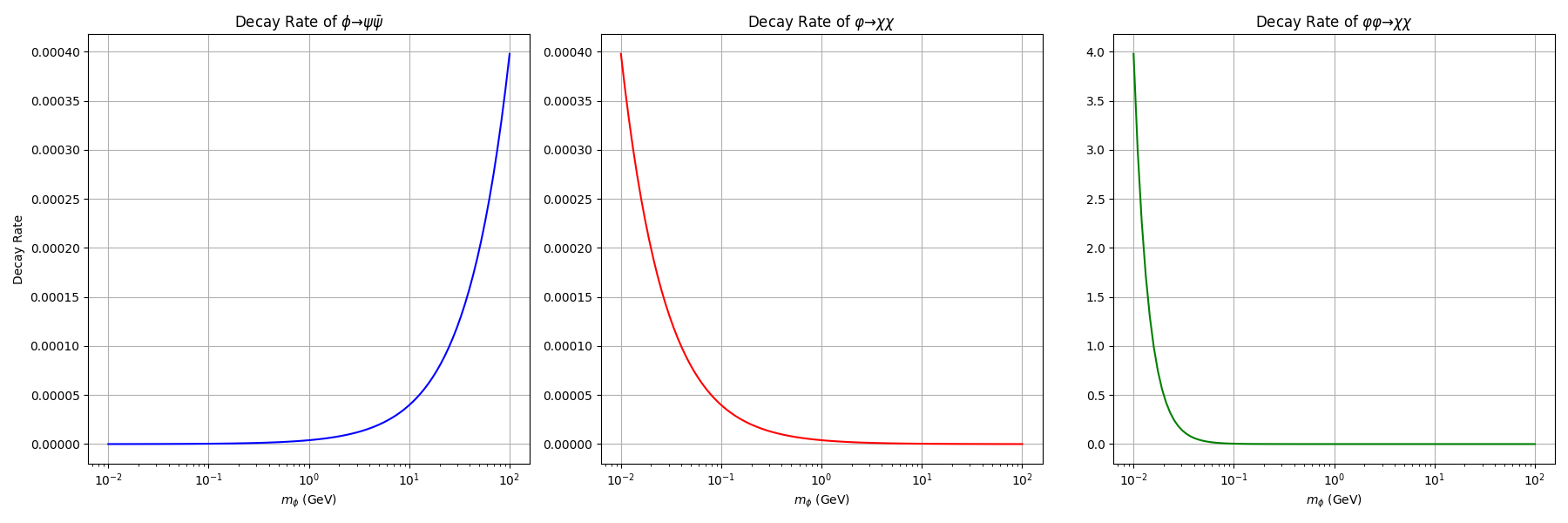}
\caption{{} The Decay Rates of the Inflaton Scalar Field with Varying Mass 
$m_{\protect\phi}$.}
\label{fig:2}
\end{figure}

Fig. \ref{fig:2} illustrates the decay rates of the inflaton $\left( \phi
\right) $\ into different particles as a function of its mass $\left(
m_{\phi }\right) $\ on a logarithmic scale, with three distinct decay
processes: $\phi \rightarrow \psi \psi ,\phi \rightarrow \chi \chi $, and\ $%
\phi \phi \rightarrow \chi \chi $. In the first subplot, the decay rate of
the inflaton into fermions\ $\left( \psi \right) $ increases rapidly for $%
m_{\phi }\geq 10GeV$, indicating that higher inflaton mass decay more
efficiently into fermions. The second subplot shows that the decay rate into
bosons $\left( \chi \right) $\ decreases as $\left( m_{\phi }\right) $\
increases, suggesting reduced efficiency for higher $m_{\phi }$ decaying
into bosons. The third subplot presents the annihilation process where two
inflatons decay into bosons, with a higher decay rate for $m_{\phi
}<10^{-1}GeV$.\ The behavior of these decay rates underscores the importance
of inflaton mass and effective couplings in determining the efficiency of
various decay channels. Higher inflaton masses tend to decay more
efficiently into fermions and through annihilation processes, while decay
into bosons is less favored at higher masses. These insights are crucial for
understanding the reheating phase post-inflation, as they influence the
universe's thermal history, particle production, and its subsequent
evolution.

\section{\label{sec:5}Preheating Dynamics in the Early Universe}

This section provides an overview of the conditions under which the issue of
thermalization arises following inflation. The inflationary model entails
the rapid expansion of the early universe, driven by a vacuum-like equation
of state. This type of equation of state can be achieved through various
means, often involving the uniform condensation of one or more classical
scalar fields. In this discussion, we will examine a model involving multiple scalar fields \cite{K4,G2}. Our focus will be on the decay of the homogeneous inflaton condensate into
inhomogeneous modes of either the same or other scalar fields, as well as
the resulting interactions between these inhomogeneous modes as they
approach thermal equilibrium. Any particles present before or during
inflation become diluted during the rapid expansion, so at the end of
inflation, all energy is stored in the potential of one or more classical,
slowly-moving, uniform inflaton fields. Immediately after inflation, the
background field(s) move rapidly and generate particles from the coupled
fields. These newly created particles interact with each other and must
eventually reach a state of thermal equilibrium. However, these particles
may be generated at such a high rate that they remain in non-equilibrium
states with exceedingly high occupation numbers for a period of time. In the
begining, lets consider chaotic inflation coupled to the a $\chi -$field 
\cite{K8}%
\begin{equation}
V(\phi ,\chi )=\frac{m_{\phi }^{2}}{2}\phi ^{2}+\frac{m_{\chi }^{2}}{2}\chi
^{2}+\frac{1}{2}g^{2}\phi ^{2}\chi ^{2}.  \label{eq27}
\end{equation}

In our previous investigation, we assumed that the decay probability $\Gamma 
$ of the scalar field $\phi $ could be calculated using conventional quantum
field theory techniques, specifically for the decay process $\phi
\rightarrow \chi \chi $, which implies that the inflaton $\phi $ has a
four-leg coupling to the scalar field $\chi .$ For simplification, we focus
on the interaction between the classical inflaton field $\phi $ and the
quantum scalar field $\chi $, utilizing the Heisenberg representation for $%
\chi $,%
\begin{equation}
\chi (\mathbf{x},t)=\frac{1}{\left( 2\pi \right) ^{3/2}}\int d^{3}k\left( 
\hat{a}_{k}\chi _{k}(t)e^{-i\mathbf{kx}}+\hat{a}_{k}^{+}\chi _{k}^{\ast
}(t)e^{i\mathbf{kx}}\right) ,  \label{eq28}
\end{equation}%
here $\hat{a}_{k}$ and $\hat{a}_{k}^{+}$ represent the annihilation and
creation operators, respectively. The equation of motion for the $\chi -$%
field is derived as follows \cite{K2}: 
\begin{equation}
\ddot{\chi}_{k}+3H\dot{\chi}_{k}+\left( \frac{\mathbf{k}^{2}}{a^{2}}+m_{\chi
}^{2}+g^{2}\phi ^{2}\right) \chi _{k}=0.  \label{eq29}
\end{equation}%
The equation describes an oscillator with a periodically varying frequency
given by $\omega _{k}^{2}=\mathbf{k}^{2}/a^{2}+m_{\chi }^{2}+g^{2}\Phi
^{2}\sin \left( mt\right) $. Due to this periodicity, modes with specific
values of $k$ can undergo parametric resonance. A straightforward way to
describe this critical phenomenon is by making a variable substitution of $%
mt=2z-\pi /2$, which simplifies the equation into the well-known Mathieu
equation \cite{K9}:%
\begin{equation}
\chi _{k}^{\prime \prime }+\left( A_{k}-2q\cos ~2z\right) \chi _{k}=0.
\label{eq32}
\end{equation}%
A key characteristic of the solution to Mathieu's equation is the presence
of an exponential instability, leading to the exponential growth of the
occupation numbers of quantum fluctuations. This phenomenon can be
interpreted as particle production \cite{K2}. Several methods can be
utilized to illustrate the behavior of scalar fields, each offering benefits
for studying different phenomena. The primary data includes the field's
value $\chi (\mathbf{x},t)$, or its Fourier transform $\chi _{k}(t)$ as an
equivalent. One of the simplest quantities that can be computed from these
values is the variance \cite{K8},%
\begin{equation}
\left\langle \left( \chi (t)-\bar{\chi}(t)\right) ^{2}\right\rangle =\frac{1%
}{\left( 2\pi \right) ^{3}}\int d^{3}k\left\vert \chi _{k}(t)\right\vert
^{2},  \label{eq33}
\end{equation}%
here, the mean value is denoted as $\bar{\chi}.\ $One of the most intriguing
quantities to evaluate is the (comoving) number density of particles in the $%
\chi $-field is given as follows \cite{K8}, 
\begin{equation}
n_{\chi }(t)\equiv \frac{1}{\left( 2\pi \right) ^{3}}\int d^{3}kn_{k}(t).
\label{eq34}
\end{equation}%
From Eq. (\ref{eq32}), as the modes $\chi _{k}$ expand, the occupation
numbers of the created particles $n_{k}(t)$ increase. The number density $%
n_{\chi }$ of particles with momentum $\mathbf{k}$ can be determined by
evaluating the energy of the mode divided by the energy $\omega _{k}$\ \cite%
{K2},%
\begin{equation}
n_{k}(t)\equiv \frac{1}{2\omega _{k}}\left\vert \dot{\chi}_{k}\right\vert
^{2}+\frac{\omega _{k}}{2}\left\vert \chi _{k}\right\vert ^{2}.  \label{eq35}
\end{equation}%
As previously mentioned, once inflation ends, the universe becomes cold and
the reheating process subsequently heats the universe to the temperatures
necessary for Big Bang nucleosynthesis. Typically, preheating is
characterized by a rapid growth of particle production that occurs at the
onset of the reheating phase. Recent studies in \cite{K6,K7} have
demonstrated that its duration can be quantified in terms of the number of
e-folds $N_{pre}$, which can be analyzed using the following equation,%
\begin{eqnarray}
N_{pre}+\frac{1-3\omega }{4}N_{re} &=&\left[ 61.1-\frac{1}{4}\ln \left( 
\frac{V_{end}}{\gamma H_{k}^{4}}\right) -N_{k}\right] ,  \label{Npre} \\
N_{re} &=&\frac{1}{3\left( 1+\omega \right) }\ln \left( \frac{3^{2}\cdot
5V_{end}}{\gamma \pi ^{2}g_{\ast }T_{re}^{4}}\right) ,
\end{eqnarray}%
here $\gamma $ is the ratio of the energy density at the end of inflation to
the preheating energy density which takes values around $\gamma = [10^{3},10^{5}]$, $\omega $\ is the equation of state that takes values in
the interval $\left[ -1/3,1/4\right] $, $H_{k},$\ $N_{k}$ and $V_{end}$\ are
inflationary parameters which can be calculated considering the potential in
Eq. (\ref{eq27}) assuming that during and at the end of inflation it was
negligibly coupled to the $\chi $\ field. $T_{re}$\ is the reheating
thermalization temperature. Eq. (\ref{Npre}) is a valuable method to
constrain preheating according to recent observations by Planck satellite 
\cite{K5}. Now we have to compute the inflationary parameters as functions
of the spectral index and $A_{s}$ which were constrained from recent
observations as $n_{s}=0.9649\pm 0.0042$\ and $A_{s}=2.196\times 10^{-9}$,\
we first define the inflationary e-fold in the following $N_{k}\simeq \frac{1%
}{M_{p}^{2}}\int_{\phi _{e}}^{\phi _{k}}\frac{V}{V^{\prime }}d\phi ,$ while $%
H_{k}$\ can be calculated as functions of $n_{s}$\ following these steps.\
Using the definition of the tensor-to-scalar ratio\ $r=2H_{k}^{2}/\pi
^{2}M_{p}^{2}A_{s},$\ Then using $r=16$ gives $H_{k}\simeq \pi M_{p}\sqrt{%
8A_{s}\epsilon }.$\ Now, for the case of our chosen potential, one computes $%
V_{end}$ in terms of $n_{s}$ and $A_{s}$, using $\epsilon (\phi _{end})=1,$
then we compute the parameters $N_{k}$ and $H_{k}$ as,%
\begin{eqnarray}
N_{k} &=&\frac{2}{\left( 1-n_{s}\right) }, \\
H_{k} &=&\pi M_{p}\sqrt{\pi A_{s}\left( 1-n_{s}\right) }.
\end{eqnarray}

\begin{figure}[tbp]
\centering
\includegraphics[width=14cm]{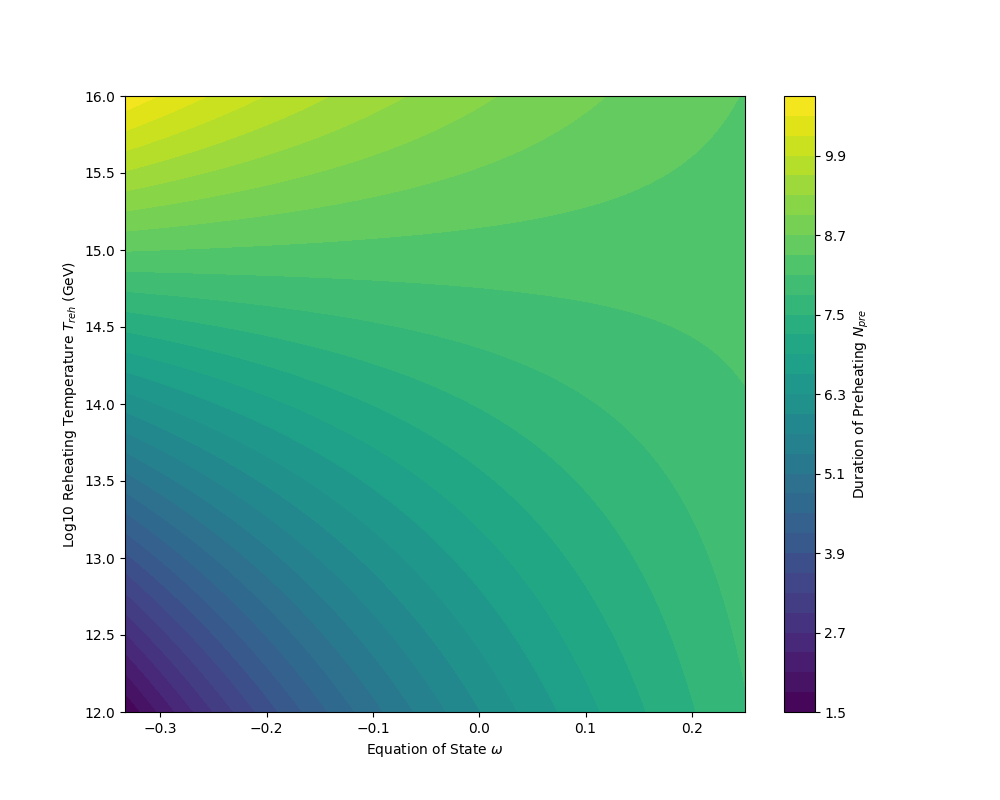}
\caption{{} Duration of Preheating as a Function of $\protect\omega$ and $%
T_{re}$}
\label{fig:3}
\end{figure}

In Fig. \ref{fig:3}\ the plot represents the duration of preheating $\left(
N_{pre}\right) $\ as a function of the equation of state parameter\ $\left(
\omega \right) $\ and the logarithm of the reheating temperature $\left(
T_{re}\right) $. The x-axis shows $\omega ,$\ which varies from $-1/3$\ to $%
1/4$\ , reflecting the relationship between pressure and energy density
during reheating. The y-axis displays the logarithm of the reheating
temperature, spanning $10^{12}$\ to $10^{16}GeV$, indicating the energy
scale at which reheating occurs. The color gradient on the plot represents
the preheating duration, with darker regions indicating shorter durations
and lighter regions indicating longer durations. From the plot we include
the finding that lower values of $\omega <0$, result in lower preheating
durations for $T_{re}<10^{15}GeV$, while the preheating duration became $%
N_{pre}>7$ for $T_{re}>10^{15}GeV$ with $\omega =[-1/3,1/4]$. Conversely, as 
$\omega $ increases towards positive values, indicating a transition towards
radiation-like behavior, the duration of preheating increases. Overall, the
plot provides valuable insights into how the equation of state and reheating
temperature influence the duration of the preheating phase.
\begin{figure}[tbp]
\centering
\includegraphics[width=14cm]{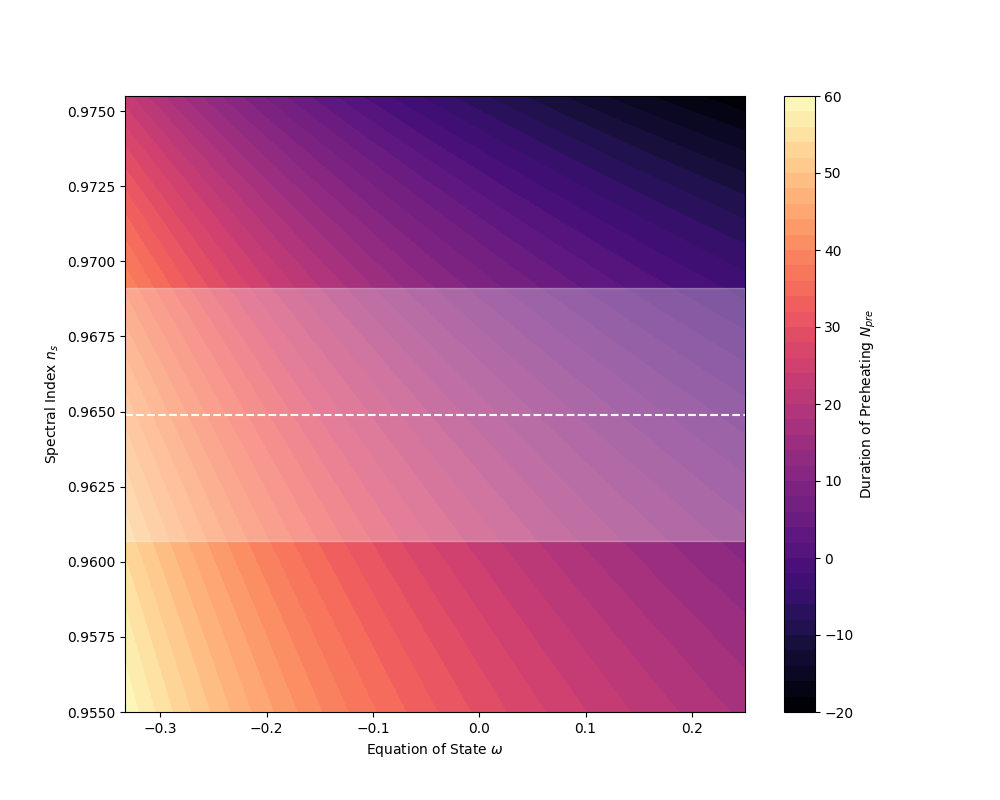}
\caption{{}  Preheating Duration as a function of the equation of state $\omega$  and the spectral index $n_{s}$}
\label{fig:4}
\end{figure}

In Fig. \ref{fig:4} the contour heatmap illustrates the duration of
preheating as a function of the equation of state parameter and the spectral
index $n_{s}.$\ The x-axis represents $\omega $, while the y-axis displays
the spectral index, which measures the density fluctuations in the early
universe. The color gradient represents the preheating duration, with darker
shades corresponding to shorter durations and lighter shades indicating
longer durations. The plot highlights observational bounds for\ $n_{s}$. The
spectral index also impacts preheating duration, with the observational
bounds being critical for aligning theoretical predictions with observed
data. Notably, negative values of $N_{pre}$\ are not considered, as they
lack physical interpretation and do not correspond to a feasible preheating
duration. Moreover, for the preheating duration to be aligned with the
observational bound the duration must be around $N_{pre}\sim \lbrack 0,40]$.

\section{\label{sec:6}Insights into Cosmological Signatures}

\subsection{The Curvature Perturbation}

\subsubsection{Curvature Fluctuation in Oscillating Inflation}

Our present focus is on analyzing cosmological perturbations in oscillating
inflation, excluding interactions with other scalar fields that lead to
inflaton decay during reheating. The useful and well-behaved perturbed
quantity during the oscillating phase has been identified as \cite%
{K10,K11,K12},%
\begin{equation}
Q=\delta \phi -\frac{\dot{\phi}}{H}\zeta .
\end{equation}%
Here, the perturbations of the scalar field and spatial curvature are
represented by $\delta \phi $ and $\zeta $, respectively \cite{K13}. The
variable $Q$, which is gauge-invariant was introduced by Mukhanov \cite%
{L4,L5}, plays a role in this context. From the evolution equation, by neglecting the third term inside the
bracket, the equation is simplified into the perturbation equation $\delta
\phi $, which disregards the metric perturbation. Consequently, we
understand that the influence of gravitational interaction on the scalar
field perturbation is encapsulated in the third term inside the bracket \cite%
{L6},%
\begin{equation}
\zeta _{c}=a+b\int \frac{dt}{a^{3}}\frac{H^{2}}{\dot{\phi}^{2}},
\end{equation}%
where $\zeta _{c}=HQ/\dot{\phi},$ $a$ and $b$ are the integration constants.
The solution proportional to the coefficient $a$ is termed the growing mode,
whereas the one proportional to $b$ is known as the decaying mode \cite{L6}.
The Bardeen parameter \cite{K12}, also referred to as the curvature
perturbation, is closely associated with the temperature fluctuation $\left(
\Delta T/T\right) $ of the Cosmic Microwave Background observed by the
recent observations \cite{K5}. We can express the observable power spectrum
of the curvature perturbation $P_{\zeta }(k)$ by equating the amplitude of
the quantum fluctuation $Q$ generated during oscillating inflation inside
the horizon as follows:%
\begin{equation}
P_{\zeta }(k)\equiv \frac{k^{3}}{2\pi ^{2}}\left\langle \left\vert \zeta
_{c}\right\vert ^{2}\right\rangle .
\end{equation}

The spectrum provides insights into the nature of the perturbations
generated during inflation. The amplitude of the power spectrum at different 
$k$ values reflects the strength of the perturbations at those scales,
knowing that higher amplitudes indicating stronger perturbations. The
behavior of the power spectrum at the extremes of $k$ can offer information
about the physics of the early universe. For example. The curvature
perturbation is directly related to the temperature fluctuations observed in
the CMB. Consequently, the power spectrum $P_{\zeta }(k)$ has a direct
connection to the observable anisotropies in the CMB making this equation
highly relevant for understanding the early universe's conditions.

\subsubsection{Curvature Power Spectrum Resulting from Preheating}

A perturbation is considered adiabatic when the fractional change $\delta x/%
\dot{x}$, is identical for all perturbations $\delta x$, where $\dot{x}$
represents the time-dependent background value. In a universe dominated by a
single fluid with a known equation of state, or by a single scalar field
with perturbations originating from the vacuum state where only adiabatic
perturbations can occur. While the adiabatic condition is not exclusive to
systems with more than one fluid, it remains preserved in certain scenarios,
such as when a single inflaton field decays into multiple components.
However, perturbations in a secondary field, such as the field into which
the inflaton decays during preheating, typically violate the adiabatic
condition \cite{L7}.

The evolution of $\zeta $ in linear theory is well-understood and arises
from the non-adiabatic component of pressure perturbations. The pressure
perturbation can be decomposed into adiabatic and non-adiabatic components
in any gauge by expressing it as follows \cite{L8}:

\begin{eqnarray}
\delta p &=&\frac{\dot{p}}{\dot{\rho}}\delta \rho +\delta p_{nad}, \\
\delta p_{nad} &=&\dot{p}\left( \frac{\delta p}{\dot{p}}-\frac{\delta \rho }{%
\dot{\rho}}\right) .
\end{eqnarray}%
Since the perturbations in the other field are not correlated with those in
the inflaton field, they do not meet the adiabatic condition. As a result,
the curvature perturbation $\zeta $ may change on large scales. To assess
the significance of this effect, a direct calculation is necessary. In this
study, we will compute this effect using the simplest preheating model described as \cite{D10,D11},%
\begin{equation}
V(\phi ,\chi )=\frac{1}{2}m^{2}\phi ^{2}+\frac{1}{2}g^{2}\phi ^{2}\chi ^{2}.
\end{equation}%
where $g$ is the coupling constant between the inflaton and the $\chi $
field. After inflation, amplified quantum fluctuations in the $\chi $ field
follow the wave equation \cite{D11},%
\begin{equation}
\delta \ddot{\chi}+3H\delta \dot{\chi}+\left( \frac{k^{2}}{a^{2}}+g^{2}\phi
^{2}\right) \delta \chi =0,  \label{EoMx}
\end{equation}%
where $k$\ is the comoving wavenumber and $a$ is the scale factor.\ The $%
\chi $ field with its effective mass leads to efficient preheating,
characterized by large amplitude oscillations when $q\equiv g^{2}\Phi
^{2}/4m^{2}\gg 1$. The growth of $\chi $ field fluctuations during
preheating gives rise to the non-adiabatic curvature perturbation $\zeta
_{nad}$. Since the inflaton decay can violate the adiabatic condition during
preheating, pressure perturbations can be divided into adiabatic and
non-adiabatic components. The evolution of $\zeta $ arises from the
non-adiabatic part of these pressure perturbations. Additionally,
non-adiabatic perturbations can cause a change in $\zeta $ on arbitrarily
large scales when these pressure perturbations are significant. In fact,
variations in $\zeta $ during preheating could be driven by the
non-adiabatic component of the $\chi $ field perturbation. The power
spectrum resulting from this amplification is given by \cite{D11},

\begin{equation}
{\mathcal{P}}_{\zeta _{nad}}\simeq \frac{2^{9/2}3}{\pi ^{5}\mu ^{2}}\left( 
\frac{\Phi }{M_{P}}\right) ^{2}\left( \frac{H_{end}}{m}\right) ^{4}\frac{%
g^{4}}{q^{\frac{1}{4}}}\left( \frac{k}{k_{end}}\right) ^{3}I(\kappa ,m\Delta
t),  \label{Pc}
\end{equation}%
here $\kappa $ is defined $\kappa ^{2}\equiv \frac{1}{18\sqrt{q}}\left( 
\frac{k}{k_{end}}\right) $, 
\begin{equation}
I(\kappa ,m\Delta t)\equiv \frac{3}{2}\int_{0}^{\kappa _{cut}}d\kappa
^{\prime }\int_{0}^{\pi }d\theta e^{2(\mu _{\kappa ^{\prime }}+\mu _{\kappa
-\kappa ^{\prime }})m\Delta t}\kappa ^{\prime 2}\sin (\theta ),  \label{0}
\end{equation}%
here $\theta $ is the angle between $\kappa ^{\prime }$ and $\kappa $, $%
\kappa _{cut}$ is an ultraviolet cut-off. The comoving wavenumber at the
Hubble radius exit and the end of inflation are denoted by $k$\ and $k_{end}$%
, respectively. The term $m\Delta t$ is an alternative way to estimate how
long the process of preheating will proceed, while $\mu $ is chosen to be $%
\mu =(\ln 3)/2\pi $ for $q\gg 1$. At the end of inflation when $m\Delta t=0$%
, the integral appeared in the power spectrum ${\mathcal{P}}_{\zeta _{nad}}$
is estimated to $I(\kappa ,0)=\kappa _{cut}^{3}\sim 1$ \cite{D11}.

According to \cite{D11} at later times of preheating, the integral appearing
in Eq. (\ref{0}) is calculated as%
\begin{equation}
I(\kappa ,m\Delta t)=0.86(m\Delta t)^{-3/2}e^{4\mu m\Delta t},
\end{equation}%
knowing that they discovered that preheating has a negligible effect on $%
\zeta $ at the scales relevant to the formation of large-scale structures.
However, smaller-scale fluctuations with wavenumbers $k\sim k_{end}$, where $%
k_{end}$ is the wavenumber corresponding to the end of preheating, may
become significant. These smaller-scale fluctuations could lead to events
such as the formation of Primordial Black Holes, as the power spectrum $%
P_{\zeta _{nad}}$ evolves with $k^{3}$.

\begin{figure}[tbp]
\centering
\includegraphics[width=16cm]{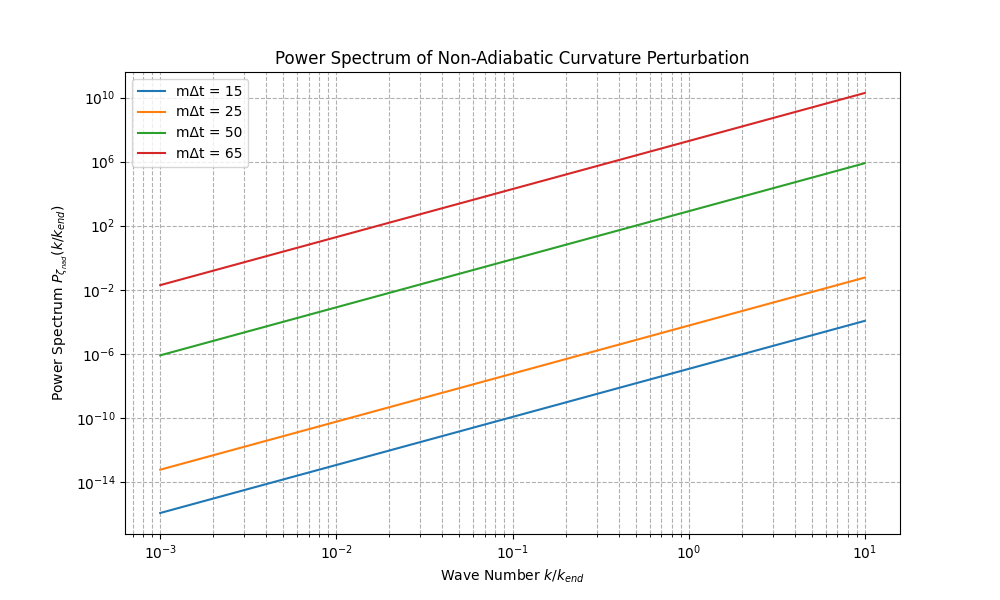}
\caption{{}The power spectrum of the non-adiabatic curvature perturbation as
a function of the normalized wave number for different values of $m\Delta t.$%
}
\label{fig:6}
\end{figure}

In Fig. \ref{fig:6} we illustrates the power spectrum $\left( P_{\zeta
_{nad}}\right) $\ of the non-adiabatic curvature perturbation as a function
of the wave number ratio $\left( k/k_{end}\right) $\ for different values of 
$m\Delta t$, which characterizes the duration of the preheating phase. The
horizontal axis represents the normalized wave number, with smaller values
corresponding to larger-scale perturbations and larger values to
smaller-scale perturbations. The vertical axis denotes the power spectrum of
the non-adiabatic curvature perturbation, quantifying the amplitude of
perturbations at different scales, the power spectrum scales with $k^{3}$.
The curves represent various $m\Delta t$\ values $(15,25,50,65)$, knowing
that shorter preheating durations resulting in lower amplitude power
spectra, suggesting weaker non-adiabatic perturbations. Conversely, longer
preheating durations show higher amplitudes, indicating significant growth
of these perturbations over time. The exponential term in\ $I(\kappa
,m\Delta t)$\ leads to rapid growth of the power spectrum with increasing $%
m\Delta t,$\ reflecting how extended preheating periods amplify
non-adiabatic perturbations.

This analysis highlights the importance of non-adiabatic perturbations
generated during preheating and their potential impact on early universe
dynamics, such as the formation of Primordial Black Holes. The sensitivity
of the power spectrum amplitude to preheating duration provides insights
into the preheating mechanism and its cosmological effects.

\subsubsection{The Abundance\ Primordial Black Holes}

Large density perturbations can cause the formation of Primordial Black
Holes in the early Universe{}{} \cite{L9}. Previous studies have
investigated various observational constraints on the abundance of PBH{}{}.
These studies indicate that the abundance of PBH is less than $10^{-20}$ of
the total energy density of the Universe. To estimate the production rate of
PBH, the fraction of energy density is calculated based on the variance of
the density perturbations. This calculation uses the power spectrum and a
window function $W(kR)$ \cite{L11} to define the variance $\sigma (k)$ \cite%
{L7},\ 
\begin{eqnarray}
\sigma ^{2}(k) &=&\frac{16}{81}\int_{0}^{\infty }\left( \frac{\tilde{k}}{k}%
\right) ^{4}P_{\zeta _{nad}}(\tilde{k})W(\tilde{k}R)\frac{d\tilde{k}}{\tilde{%
k}}, \\
\sigma ^{2}(k) &\approx &10\sqrt{2\pi }\frac{2^{9/2}3}{\pi ^{5}\mu ^{2}}%
\left( \frac{\Phi }{M_{p}}\right) ^{2}\left( \frac{H_{end}}{m}\right)
^{4}g^{4}q^{1/2}\left( \frac{k}{k_{end}}\right) ^{3}.
\end{eqnarray}

Assuming that the primordial curvature perturbations follow Gaussian
distributions, we can estimate the abundance of PBH \cite{L11}. The fraction
of the energy density that collapses into PBH can be approximated as
follows{} \cite{L12,L13}:%
\begin{equation}
\beta (k)\simeq \sqrt{\frac{1}{2\pi }}\frac{\sigma }{\delta _{c}}\exp \left(
-\left( \frac{\delta _{c}}{\sqrt{2}\sigma (k)}\right) ^{2}\right) .
\end{equation}

Here, $\beta (k)$\ represents the fraction of energy density that collapses
into PBH at a given scale $k$\ and, $\sigma \left( k\right) $\ is the
standard deviation of the density perturbations at the scale $k$, and $%
\delta _{c}$\ is the critical density threshold for collapse. This
expression indicates that the probability of PBH formation is exponentially
sensitive to the ratio of the critical density threshold to the standard
deviation of perturbations{}{}. Observational constraints on the abundance
of PBH are critical for understanding their role in the Universe. Several
methods have been employed to constrain $\beta (k)$. For instance, the CMB
observations can limit the amount of PBH through their effects on CMB
anisotropies. Gravitational lensing surveys can detect or constrain PBH by
observing the lensing effects they produce on background objects.
Furthermore, the distribution and dynamics of large-scale structures can
provide upper limits on PBH abundance, as an excessive number of PBH would
disrupt the observed structure formation. These constraints are crucial as
they help to refine theoretical models and simulations of PBH formation.
Future observations, particularly those aimed at detecting gravitational
waves from PBH mergers or more detailed lensing surveys, may offer more
stringent constraints or even direct evidence of PBH. Such observational
efforts will enhance our understanding of the early Universe and the
potential role of PBH in cosmic evolution{}{}. These constraints are crucial
as they help to construct theoretical models and simulations of PBH
formation.

\begin{figure}[tbp]
\centering
\includegraphics[width=16cm]{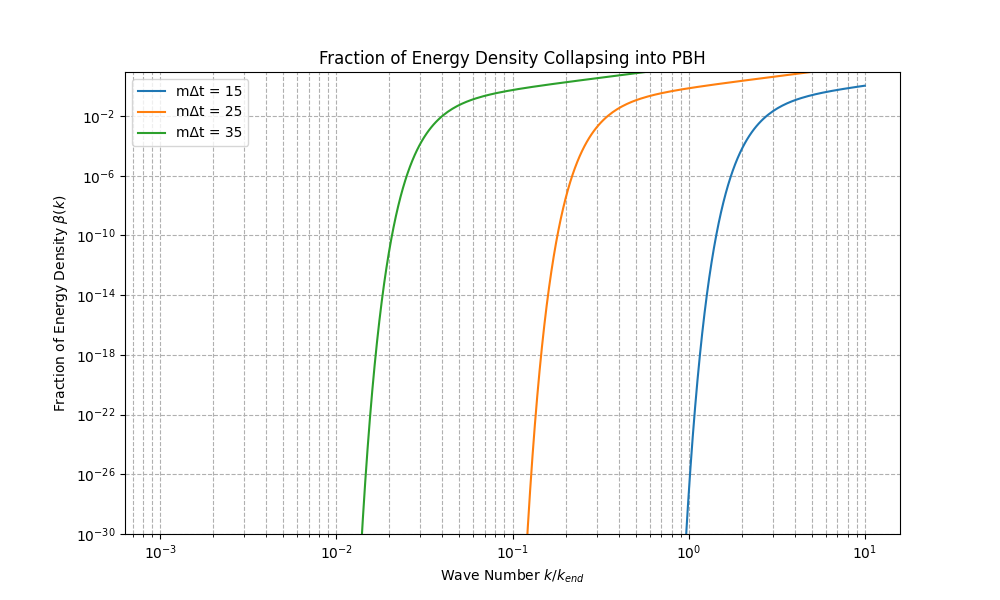}
\caption{{}Fraction of Energy Density Collapsing into PBH}
\label{fig:7}
\end{figure}

One important question is whether the abundance of PBHs and their effect
through gravitational waves can be detected through observational methods.
For this reason, the fraction of the energy density can give an alternative
explanation to study PBH collapses in the early Universe. Fig. (\ref{fig:7})
shows the fraction of energy density, $\beta (k)$, that collapses into
primordial black holes as a function of the fraction of the wave number\ $k$%
\ and $k_{end}$. The curves represent different values of the product\ $%
m\Delta t$, indicating various stages in the time evolution after inflation.
As the wave number increases, generally increases up to a certain point,
showing that PBH formation is scale-dependent. Different $m\Delta t$\ values
shift the curves, highlighting the impact of time evolution on PBH
formation. For instance, at $m\Delta t=15,$\ the fraction of energy density
collapsing into PBH is lower compared to $m\Delta t=25$\ and $m\Delta t=35$.
The peaks in the curves indicate the most probable scales for PBH formation
at different times, which shift with changing $m\Delta t$. This plot shows how primordial perturbations at various scales
contribute to PBH formation and how this formation evolves with time after
the end of inflation. It helps constrain the parameters governing PBH
formation and informs observational strategies for detecting PBH.

\subsubsection{Primordial Gravitational Waves}

Gravitational waves are ripples in the fabric of spacetime that are generated by the
acceleration of massive objects and are characterized by their
transverse-traceless nature, meaning they do not have longitudinal or trace
components. During the preheating phase, violent interactions between the
inflaton field and other matter fields can efficiently produce gravitational
waves. The evolution of these waves is governed by the linearized Einstein
equations, and in the context of the FRW background, they manifest as
perturbations to the spatial part of the metric, specifically the traceless
part. This approach allows us to isolate the gravitational wave
contributions from other scalar and vector perturbations, providing a
clearer understanding of the early Universe's dynamics and the mechanisms
that lead to the production of gravitational radiation. The post-inflation
production of matter fields can cause significant changes in the metric. Our
focus is oriented to the evolution of the transverse-traceless metric
perturbation $h_{ij}$ and the generation of gravitational waves during
preheating. In the FRW background, gravitational waves can be expressed as
the traceless component of the spatial metric perturbations \cite{L14},%
\begin{eqnarray}
ds^{2} &=&g_{\mu \nu }dx^{\mu }dx^{\nu },  \notag \\
&=&-dt^{2}+a(t)^{2}\left( \delta _{ij}+h_{ij}\right) dx^{i}dx^{j}.
\end{eqnarray}%
The equation of motion for the perturbation $h_{\mu \nu }$, which satisfies
the transverse-traceless (TT) conditions $\partial _{i}h_{ij}=h_{ii}=0$, can
be written as \cite{L15}:%
\begin{equation}
\ddot{h}_{ij}+3H\dot{h}_{ij}-\frac{1}{a^{2}}\nabla ^{2}h_{ij}=2\kappa
^{2}\Pi _{ij}^{TT},
\end{equation}%
here $\kappa ^{2}=1/M_{p}^{2}$, the source term $\Pi _{ij}^{TT}$ is the
transverse-traceless of the anisotropic stress $\Pi _{ij}.$The energy
density associated with gravitational waves can be computed using the
following equation \cite{L16}, 
\begin{equation}
\rho _{GW}=\frac{1}{4\kappa ^{2}}\left\langle \dot{h}_{ij}(t,\mathbf{x})\dot{%
h}_{ij}(t,\mathbf{x})\right\rangle .
\end{equation}

The energy spectrum of GWs represents their abundance of energy density at
present, and is a measure of their strength. It can be expressed as:%
\begin{equation}
h^{2}\Omega _{GW,0}(f)=\frac{h^{2}}{\rho _{c,0}}\frac{d\rho _{GW,0}}{d\ln f},
\end{equation}%
where $f$ is the frequency and $\rho _{GW,0}$\ is the critical energy
density today. Since we are interested in correlating the gravity-wave
energy density spectrum with current observations, we must translate the
previously derived GW spectrum into physical quantities. The present scale
factor, compared to the scale factor when GW production ceases, can be
expressed as \cite{L17}:

\begin{equation}
\frac{a_{end}}{a_{0}}=\frac{a_{end}}{a_{pre}}\left( \frac{a_{pre}}{a_{re}}%
\right) ^{1-\frac{3}{4}\left( 1+\omega \right) }\left( \frac{\bar{g}_{\ast }%
}{\bar{g}_{0}}\right) ^{\frac{-1}{12}}\left( \frac{\rho _{r,0}}{\rho _{\ast }%
}\right) ^{\frac{1}{4}}.
\end{equation}

Supposing that GW production stops at the end of preheating, let $"_{pre}"$
represent the time when GW production is finished, and $"_{re}"$ represent
the time when thermal equilibrium is reached. Here, $\rho _{r,0}$\ denotes
the present radiation energy density, and the total energy density of the
scalar field is represented by $\rho _{\ast }.$\ Knowing that $\frac{\bar{g}%
_{\ast }}{\bar{g}_{0}}\simeq 31.$The corresponding physical frequency today
is given by \cite{L18},%
\begin{equation}
f=\frac{k}{2\pi a_{0}}=\frac{k_{0}}{a_{end}\rho _{\ast }^{1/4}}\times \left(
4\times 10^{10}Hz\right) \ .
\end{equation}%
Knowing that the abundance of radiation today given as\ $\Omega
_{r,0}h^{2}=h^{2}\rho _{r,0}/\rho _{c,0},$ with $h,$\ is the current
dimensionless Hubble constant and $\Omega _{GW,0}h^{2}$ \cite{L17}. In Refs. 
\cite{K6,K7} \ a valuable result was proven in which they directly related
the energy spectrum of GWs to the preheating duration which is an
alternative method to study the constraints on the gravitational wave
produced during the stage of preheating, the equation is written in the
following way,%
\begin{equation}
\Omega _{GW}(f)=\frac{\Omega _{GW,0}h^{2}}{\Omega _{r,0}h^{2}}\left( \frac{%
g_{\ast }}{g_{0}}\right) ^{1/3}e^{4N_{pre}},
\end{equation}%
here it has been shown the possibility to set a correlation between
gravitational wave energy density spectrum and the observable duration of
preheating according to recent Planck Data.

\begin{table*}[tbp]
\centering%
\begin{tabular}{c|c|c|c|c|c|c||c}
\hline
$n_{s} $ & $\omega $ & $H_{k}[GeV] $ & $T_{re}[GeV] $ & $N_{k} $ & $N_{pre} $
& $\Omega_{GW,0}h^{2} $ & $\Omega_{GW}f $ \\ \hline\hline
$0.9646 $ & $-0.014 $ & $1.145\cdot 10^{14} $ & $6.014\cdot 10^{11} $ & 54.72
& 0.046 & 1.1837$\cdot 10^{-6} $ & 6.192$\cdot 10^{-2} $ \\ 
$0.9649 $ & $-0.0451 $ & $1.192\cdot 10^{14} $ & $5.706\cdot 10^{11} $ & 
54.27 & 0.528 & 1.2151$\cdot 10^{-6} $ & 8.385$\cdot 10^{-1} $ \\ 
$0.9653 $ & $-0.059 $ & $1.190\cdot 10^{14} $ & $1.389\cdot 10^{11} $ & 54.48
& 0.219 & 1.2752$\cdot 10^{-6} $ & 7.262$\cdot 10^{-2} $ \\ 
$0.9638 $ & $-0.011 $ & $1.191\cdot 10^{14} $ & $4.082\cdot 10^{11} $ & 54.35
& 0.236 & 2.8514$\cdot 10^{-7} $ & 6.933$\cdot 10^{-2} $ \\ 
$0.9648 $ & $-0.088 $ & $1.196\cdot 10^{14} $ & $1.691\cdot 10^{11} $ & 53.93
& 0.265 & 6.8902$\cdot 10^{-7} $ & 1.005$\cdot 10^{-2} $ \\ 
$0.9646 $ & $0.095 $ & $1.193\cdot 10^{14} $ & $4.536\cdot 10^{11} $ & 54.21
& 0.646 & 3.4453$\cdot 10^{-7} $ & 12.846$\cdot 10^{-1} $ \\ 
$0.9632 $ & $0.017 $ & $1.188\cdot 10^{14} $ & $7.029\cdot 10^{11} $ & 54.62
& 0.165 & 1.4056$\cdot 10^{-6} $ & 1.190$\cdot 10^{-1} $ \\ 
$0.9633 $ & $0.011 $ & $1.192\cdot 10^{14} $ & $5.229\cdot 10^{11} $ & 54.32
& 0.485 & 1.0650$\cdot 10^{-6} $ & 5.839$\cdot 10^{-1} $ \\ \hline
\end{tabular}%
\caption{Testing the density of gravitational waves preoduced during
preheating as functions of several cosmological parameters.}
\label{table:1}
\end{table*}

Table \ref{table:1} presents data related to the physical parameters of our
chosen cosmological model, particularly focusing on the early universe and
the preheating period after inflation. The preheating duration, plays a
crucial role in determining the density of gravitational waves during the
preheating phase. During preheating, energy is transferred from the inflaton
field to other fields, potentially generating significant gravitational
waves. A longer preheating duration allows more energy to be converted into
gravitational waves, increasing the density $\Omega _{GW}(f)$. This enhanced
gravitational wave production during extended preheating which can leave a
detectable imprint on the gravitational wave background, providing a unique
probe into the dynamics of the early universe. Overall, the table shows
slightly different cosmological parameters, each affecting the early
universe's dynamics and present-day observables, providing a detailed look
at how initial conditions influence the of evolution universe.\ 

\section{\label{sec:7}Conclusion}

The capability to reheat the Universe following a phase of exponential
expansion is a crucial characteristic of any inflationary model. In some
instances, it is enough to be aware of the temperature $T_{re}$ at which the
Universe reheats and reaches the equilibrium, resulting in a phase of
radiation-dominated expansion. We have considered several important aspects
of the preheating, reheating, and inflaton oscillation processes. we
discussed a basic model comprising a massive inflaton field $\phi $
interacting with a scalar field $\chi $. Despite the simplicity of the
model, the preheating theory is the best candidate to explain this
phenomenon, as demonstrated. The primary objective was not to resolve all
inquiries surrounding preheating theory but to establish a suitable
framework for conducting further research on the subject. Initially, there
is particle generation under broad parametric resonance conditions, which
subsequently narrows overtime before coming to an end. As thoroughly
discussed, large-scale curvature perturbations can fluctuate if there exists
a considerable non-adiabatic pressure perturbation. This can always occur in
theory if there are multiple fields or fluids present. Given that when
preheating involves the inflaton decays through resonance, such fluctuations
are theoretically possible. By concentrating on the most straightforward
preheating model and the non-adiabatic pressure, it was discovered that the
primary influence arises from second-order perturbations in the preheating
field. The formation of primordial black holes in the two-field preheating
model with a quadratic inflaton potential was examined. The power spectrum
of curvature perturbations displays a $k^{3}$-spectrum, implying that
fluctuations are most prominent on small scales. If overproduced during the
subsequent radiation-dominated period, primordial black holes could be
created. Finally, we revisit the fundamentals of Primordial Gravitational
Waves and the energy density they transport. These waves were generated
during the preheating phase.

\end{document}